\documentstyle[graphicx,twocolumn,prl,aps]{revtex}

\begin{document}

\pagestyle{empty} \narrowtext
\parskip=0cm \noindent

{\bf Comment on ``Subleading Corrections to Parity Violating Pion
Photoproduction"}

\smallskip

In a recent preprint, \cite{ZPHR} Zhu {\it et al.}\ calculated a part of the
next-to-leading order (NLO) corrections to parity-violating (PV) pion
photoproduction $\overrightarrow{\gamma }p\rightarrow \pi ^{+}n$ in
heavy-baryon chiral perturbation theory (HB$\chi $PT). They claim they have
found the contribution as large as the leading-order (LO). If correct, the
process will not be a clean way to extract the longest range PV pion-nucleon
coupling constant $h_{\pi NN}^{(1)}$, as was asserted in our previous
publications \cite{CJ}. In this comment we show that there is no solid
evidence to support Zhu {\it et al.}'s claim. Moreover, we show that the
subleading parity-violating coupling $h_V$ cannot be extracted from
observables at the order of interest based on the formalism of effective
field theory.

The disagreement stems from the estimation of the coefficients of the
relevant PV operators in the chiral lagrangian \cite{KS,ZPHRx,CJ} 
\begin{eqnarray}
{\cal L}^{PV} &=&-ih_{\pi NN}^{(1)}\pi ^{+}\overline{p}n-\frac{h_{V}}{\sqrt{2%
}F_{\pi }}\overline{p}\gamma ^{\mu }nD_{\mu }\pi ^{+}  \nonumber \\
&&+i\frac{h_{A}^{(1)}+h_{A}^{(2)}}{F_{\pi }^{2}}\overline{p}\gamma ^{\mu
}\gamma _{5}p\pi ^{+}D_{\mu }\pi ^{-}  \nonumber \\
&&+i\frac{h_{A}^{(1)}-h_{A}^{(2)}}{F_{\pi }^{2}}\overline{n}\gamma ^{\mu
}\gamma _{5}n\pi ^{+}D_{\mu }\pi ^{-}  \nonumber \\
&&-ie\frac{C}{m_{N}F_{\pi }}\overline{p}\sigma ^{\mu \nu }F_{\mu \nu }n\pi
^{+}+{\rm h.c.}+\cdots \ .
\end{eqnarray}
Here the coupling constants $h_{\pi NN}^{(1)}$ and $h_{V}$ are not defined
according to the standard practice of the effective field theory. Since the
operator $\overline{p}\gamma ^{\mu }nD_{\mu }\pi ^{+}$ has dimension five,
one unit higher than the leading operator $\pi ^{+}\overline{p}n$, the
natural size of its coefficient $h_{V}/\sqrt{2}F_{\pi }$ is $h_{\pi
NN}^{(1)}/\Lambda _{\chi }$. In other words, the natural size of $h_{V}$ is $%
(\sqrt{2}F_{\pi }/\Lambda _{\chi })h_{\pi NN}^{(1)}\sim (\sqrt{2}/4\pi
)h_{\pi NN}^{(1)}$, {\it not} $h_{\pi NN}^{(1)}$ as implied in the comment
in Ref. \cite{ZPHR} following Eq. (13). Indeed, naive dimensional analyses
performed in Refs. \cite{KS,Bira} have $\left| h_{\pi NN}^{(1)}\right| \sim
5\times 10^{-7}$ (which is very close to the DDH ``best guesstimate'' \cite
{DDH}) and $\left| h_{V}\right| \sim 5\times 10^{-8}$. If these estimations
are employed, then $h_{V}$ contributes only at 10\% level according to Eq.
(13) in \cite{ZPHR}.

Is there any evidence that the relative sizes of $h_{\pi NN}^{(1)}$ and $%
h_{V}$ do not follow natural power counting? To establish this, one must
show that either $h_{\pi NN}^{(1)}$ is suppressed, or $h_{V}$ is enhanced,
relative to their natural orders of magnitude. As is well known, the
experimental data on the former is still controversial \cite{CJ}, and hence
it is natural to assume at this point that $h_{\pi NN}^{(1)}$ is of order $%
\sim 5\times 10^{-7}$. It is possible, though, that $h_{\pi NN}^{(1)}$ is
much smaller than the value suggested by naive dimensional analyses as
suggested by the $^{18}$F experiments \cite{F18}. If so $h_{\pi NN}^{(1)}$
may not be the dominant effect in PV $\overrightarrow{\gamma }p\rightarrow
\pi ^{+}n$. To determine conclusively if $h_{\pi NN}^{(1)}$ is strongly
suppressed is, in fact, one of the important motivations for our papers \cite
{CJ}.

Phenomenologically, the size of $h_{V}$ is even less certain. As far as we
know, the only constraint on $h_{V}$ comes from the electroweak radiative
corrections (through the axial current) to the single-spin asymmetry
measured by SAMPLE in PV $\vec{e}-p$ and $\vec{e}-d$ scattering \cite{SAMPLE}%
. Since the leading-order prediction strongly disagrees with current data, a
huge size radiative correction is required to reconcile the theory and
experiment. If this correction is calculated in leading order in chiral
perturbation theory, as it has been done in Refs. \cite{ZPHRx,Bira}, an
usually large size (as much as 100 times larger than their natural size) of
the couplings $h_{A}$ and/or $h_{V}$ is needed to explain the data. This,
however, is a strong indication that the power counting for the radiative
correction itself breaks down and the data cannot be used straightforwardly
to extract a reliable $h_{V}$. Given that the SAMPLE result is not yet fully
understood, we believe that the benefit of doubt should be given to a
natural size $h_{V}\sim 5\times 10^{-8}$.


There is, however, a more fundamental point: $h_{V}$ cannot be extracted, in
principle, independent of other parameters, from observables calculated to
the order of interest at present. For instance, in the leading-order chiral
perturbation calculation of the isovector nucleon anapole moment \cite
{ZPHRx,Bira}, $h_{V}$ appears together with $h_{A}^{(2)}$ in a combination $%
h_{A}^{(2)}-g_{A}h_{V}/2$. Therefore, $h_{A}^{(2)}$ and $h_{V}$ cannot be
extracted independently from the value of the anapole moment. Likewise, in
the NLO result for the spin-asymmetry $\overrightarrow{\gamma }p\rightarrow
\pi ^{+}n$, $h_{V}$ and $C$ appear in the combination $C-\frac{\kappa
_{p}-\kappa _{n}}{4\sqrt{2}}h_{V}$ \cite{ZPHR}. It is easy to show, through
field redefinitions, that the above combinations are the only ones that
appear in PV observables to the order of current interest. Therefore, $h_{V}$
has no direct physical effects and can be set to zero by hand. [The
kinematical difference between the two operators seen in ref. \cite{ZPHR} is
the result of a higher-order artifact which is partially resumed to lower
order. A direct computation of the relativistic tree diagrams confirms this.]

To prove our assertion, we use the field redefinition which is a powerful
tool to eliminate the equation-of-motion operators that generally have no
effect on physical observables. Introduce the following field
transformation: 
\begin{eqnarray}
&&p\rightarrow p-\frac{ih_{V}}{\sqrt{2}F_{\pi }}\pi ^{+}n\ ,  \nonumber \\
&&n\rightarrow n-\frac{ih_{V}}{\sqrt{2}F_{\pi }}\pi ^{-}p\ .
\end{eqnarray}
Note that the Jacobian of this transformation is unity up to some negligible 
${\cal O}$($h_{V}^{2}$) corrections. The new lagrangian becomes 
\begin{eqnarray}
{\cal L}^{PV} &=&-ih_{\pi NN}^{(1)}\pi ^{+}\overline{p}n  \nonumber \\
&&+i\frac{h_{A}^{(1)}+\overline{h}_{A}^{(2)}}{F_{\pi }^{2}}\overline{p}%
\gamma ^{\mu }\gamma _{5}p\pi ^{+}D_{\mu }\pi ^{-}  \nonumber \\
&&+i\frac{h_{A}^{(1)}-\overline{h}_{A}^{(2)}}{F_{\pi }^{2}}\overline{n}%
\gamma ^{\mu }\gamma _{5}n\pi ^{+}D_{\mu }\pi ^{-}  \nonumber \\
&&-ie\frac{\overline{C}}{m_{N}F_{\pi }}\overline{p}\sigma ^{\mu \nu }F_{\mu
\nu }n\pi ^{+}+\text{h.c.}+\cdots \ ,
\end{eqnarray}
where we have omitted terms having dependence on the $\pi ^{0}$ field as
well as those with three and more pion fields and more derivatives. The
barred parameters are defined as 
\begin{equation}
\overline{h}_{A}^{(2)}=h_{A}^{(2)}-\frac{g_{A}}{2}h_{V}\ ,\quad \overline{C}%
=C-\frac{\kappa _{p}-\kappa _{n}}{4\sqrt{2}}h_{V}\ .
\end{equation}
So only two independent combinations of $h_{A}^{(2)}$, $h_{V}$ and $C$
appear in the effective lagrangian to the order shown. We do not know at
this point whether $h_{V}$ can be eliminated entirely from the effective
lagrangian. However, what we show here is that an experimental determination
of it is extremely difficult because of the chiral suppression associated
with higher orders.

To summarize, we argue that there is no credible evidence that the relative
sizes of $h_{\pi NN}^{(1)}$ and $h_{V}$ do not follow the natural power
counting. In addition, through field redefinitions, we find that $h_{V}$
cannot be extracted at the order of interest independent of other unknown
parameters of the theory.

\smallskip

\noindent Jiunn-Wei Chen and Xiangdong Ji\newline

\vspace{-0.6cm}

\begin{quote}
{\small Department of Physics, University of Maryland, College Park, MD20742 
}
\end{quote}

\noindent 

\end{document}